\documentclass[11pt,a4paper]{article}
\usepackage{jheppub}
\usepackage{etoolbox}

\def\one{{\,\hbox{1\kern-.8mm l}}}
\newcommand{\Dslash}{\not{\hbox{\kern-4pt $D$}}}
\newcommand{\pdslash}{\not{\hbox{\kern-2pt $\partial$}}}

\newcommand{\Comment}[1]{{}}

\def\IZ{{\mathbb Z}}

\def\IR{{\mathbb R}}

\newcommand{\bc}{\begin{center}}
\newcommand{\ec}{\end{center}}
\newcommand{\ba}{\begin{array}}
\newcommand{\ea}{\end{array}}
\newcommand{\beq}{\begin{equation}}
\newcommand{\eeq}{\end{equation}}
\newcommand{\bea}{\begin{eqnarray}}
\newcommand{\eea}{\end{eqnarray}}
\newcommand{\bmx}{\begin{pmatrix}}
\newcommand{\emx}{\end{pmatrix}}

\newcommand{\be}{\begin{equation}}
\newcommand{\ee}{\end{equation}}

\newcommand{\del}{\partial}
\newcommand{\half}{{\frac{1}{2}\,}}

\newcommand{\tD}{{\tilde D}}
\newcommand{\tdel}{{\tilde \del}}

\newcommand{\tll}{{\tilde \ell}}

\newcommand{\eref}[1]{Eq.\,(\ref{#1})}

\newcommand{\tc}{{\tilde c}}

\newcommand{\tilh}{{\tilde h}}
\newcommand{\tm}{{\tilde m}}

\newcommand{\tchi}{{\tilde \chi}}

\def\IB{\relax{\rm I\kern-.18em B}}
\def\IC{{\relax\hbox{\kern.3em{\cmss I}$\kern-.4em{\rm C}$}}}
\def\ID{\relax{\rm I\kern-.18em D}}
\def\IE{\relax{\rm I\kern-.18em E}}
\def\IF{\relax{\rm I\kern-.18em F}}
\def\II{\relax{\rm I\kern-.18em I}}
\def\IZ{\relax{\sf Z\kern-.35em Z}}
\def\Id{\relax{1\kern-.32em 1}}
\def\IG{\relax\hbox{$\inbar\kern-.3em{\rm G}$}}
\def\IR{\relax{\rm I\kern-.18em R}}
\newcommand\sfrac[2]{{\textstyle\frac{#1}{#2}}}

\newcommand\shalf{{\textstyle\frac12}}

\title{Cosets of Meromorphic CFTs and Modular Differential Equations}

\author[a]{Matthias R. Gaberdiel,}
\author[b]{Harsha R. Hampapura}
\author[b]{and Sunil Mukhi}

\affiliation[a]{Institut f\"ur Theoretische Physik, ETH Z\"urich,\\
CH-8093 Z\"urich, Switzerland}\affiliation[b]{Indian Institute of Science Education and Research,\\
Homi Bhabha Rd, Pashan, Pune 411 008, India}

\emailAdd{gaberdiel@itp.phys.ethz.ch}
\emailAdd{harshahr93@gmail.com}
\emailAdd{sunil.mukhi@gmail.com}

\abstract{Some relations between families of two-character CFTs are explained using a slightly generalised coset construction,
and the underlying theories (whose existence was only conjectured based on the modular differential equation) are
constructed.
The same method also gives rise to interesting new examples of CFTs with three and four characters.}

    
\begin{document}

\maketitle

\section{Introduction}
A complete understanding of the space of two-dimensional conformal field theories is not yet available, even for 
the so-called rational conformal field theories (RCFTs). 
These theories have a finite number of characters and one can, in principle, attempt to classify them using their modular 
(and other consistency) properties. In particular, in the modular differential equation approach, RCFTs are classified 
according to the number of independent characters ($n$), and the zeroes of the Wronksian ($\ell$) on the torus 
moduli space \cite{Mathur:1988na}. These 
two non-negative integers determine the modular invariant differential equation satisfied by the characters up to a finite set of 
constants. The resulting differential equation can then be solved as a function of these parameters using a power series expansion. 
Demanding that the coefficients of the power series be non-negative integers (since they correspond to degeneracies of states) one 
obtains constraints on the possible values of the parameters. In some cases, these are sufficient to determine the central charge and the 
conformal dimensions of the primaries. This method turns out to be particularly useful for the classification of theories with a 
small number of characters. In the case $\ell=0$ one can show that there are finitely many 2-character theories \cite{Mathur:1988na}, 
almost all of which turn out to be affine theories at level $1$. For three character theories there are infinitely many candidates, 
and several infinite subsets have been identified. Using differential equations one can also deduce several generic properties of 
such CFTs \cite{Mathur:1988gt}. 

Much less is known about the $\ell > 0$ case. Indeed, essentially the only concrete result until recently was the identification of 
a set of candidate two-character theories with $\ell=2$ together with their modular transformation properties \cite{Naculich:1988xv}. 
In a recent study \cite{Hampapura:2015cea} two of the present authors analysed differential equations for two-character CFTs 
with $\ell>0$. This work successfully reproduced the above list of $\ell=2$ examples and verified the integrality of the expansion 
coefficients to high orders. It also proposed potential affine symmetry algebras for these $\ell=2$ theories, but it was already clear 
from the analysis of \cite{Hampapura:2015cea} that none of these chiral algebras is generated just by the affine currents. 
As a consequence, the existence of these theories remained somewhat conjectural. It was also noted in \cite{Hampapura:2015cea} 
that these theories exhibit intriguing parallels to the $\ell=0$ examples of \cite{Mathur:1988na}
that call for an explanation.

In this paper we explain these parallels, and in the process give a construction of the family of $\ell=2$ theories. In particular,
we show that the observed relations between the $\ell=0$ and $\ell=2$ two-character theories have a natural origin in that
each $\ell=2$ two-character theory can be described as a (generalised) coset where we divide a meromorphic (self-dual) 
conformal field theory at central charge $c=24$ by the corresponding $\ell=0$ affine subtheory (with two characters). This coset
construction is somewhat novel in the sense that the numerator is not an affine theory. Our proposal not only confirms the 
existence of the $\ell=2$ two-character CFTs, but also provides a definition for them, and explains the relations between their 
properties. A similar construction (this time involving the self-dual $\mathfrak{e}_8$ theory at level one with $c=8$)
 also explains some pairwise regularities between different $\ell=0$ theories, and the cosets of $c=24$ self-dual theories
 by three- and four-character affine theories lead to new $\ell=0$ CFTs (with three and four characters, respectively). 
 For these new theories we verify, using differential equations, that the proposed characters have indeed non-negative 
 integer coefficients in their power series expansion. 
\smallskip

The paper is organised as follows. In the following section we explain carefully the generalised coset construction where only
the denominator theory is an affine theory (but the numerator theory is in general not). In Section~3 we review briefly
the modular differential equation approach to the classification of RCFTs, and work out the relation between the
parameters $(n,\ell)$ of the denominator and the coset theory (assuming that the numerator theory is a self-dual
theory, i.e., has a single character). In Section~4 we then apply this construction to interesting examples: in Section~4.1
we explain some intriguing relation between pairs of $\ell=0$ two-character theories, while Section~4.2 deals with
the main topic, the relation between the $\ell=0$ and $\ell=2$ family of two-character theories. We also use this
method in Section~4.3 to construct apparently new classes of three- and four-character theories. Finally, there
are some brief conclusions in Section~5.


\section{A family of generalised coset constructions}

Let us begin by explaining the coset construction of two-dimensional conformal field theories in some generality, slightly extending the familiar
analysis of \cite{Goddard:1986ee}; generalisations of this kind have also been considered before in 
\cite{DM,Li}. Suppose ${\cal H}$ is a meromorphic conformal field theory, for 
example a self-dual (lattice) theory, that contains an affine symmetry algebra, but whose chiral algebra 
may not be generated just by the currents. Let us denote by ${\cal D}$ an affine subtheory of ${\cal H}$, 
associated with a semi-simple Lie algebra $\mathfrak{h}$ at positive integer level $k$.  Then we can construct 
the coset theory 
\be
{\cal C} = {\cal H } / {\cal D} \ , 
\ee
whose chiral algebra contains all chiral fields of the numerator theory ${\cal H}$ that have a trivial
OPE with any of the chiral fields of the denominator theory. In the more familiar construction of say
\cite{Goddard:1986ee}, the numerator theory will also be an affine theory, but as will become apparent 
momentarily, this is not essential for the construction to work. 

In order to show that this generalised coset leads to a consistent conformal field theory, we need to show
that it possesses a stress-energy tensor of the appropriate central
charge $c^{\cal C} = c^{\cal H} - c^{\cal D}$. In order to see this we note that since the denominator
theory is an affine theory, its stress-energy tensor is given by the Sugawara construction, involving
only the currents $J^a$ from $\mathfrak{h}$. On the other hand, since the affine theory is a subtheory
of the numerator theory, we have that both
\be
{}[L^{\cal H}_n, J^a_m] = - m J^a_{n+m} \ , \qquad \hbox{and} \qquad
{}[L^{\cal D}_n, J^a_m] = - m J^a_{n+m} \ , 
\ee
where $L^{\cal H}_n$ and $L^{\cal D}_n$ are the Virasoro modes of the numerator and denominator theory,
respectively. It thus follows that 
\be
{} [L^{\cal C}_n, J^a_m] = 0 \ , \qquad \hbox{where} \qquad
L^{\cal C}_n = L^{\cal H}_n - L^{\cal D}_n \ . 
\ee
Thus the modes $L^{\cal C}_n$ are part of the coset chiral algebra. Furthermore, since the
Virasoro generators of the denominator $L^{\cal D}_m$ are bilinears in the currents $J^a_l$, we can furthermore conclude that
\be\label{central}
{}[ L^{\cal C}_n, L^{\cal D}_m] = 0 \ . 
\ee
This is then sufficient to prove that the coset modes $L^{\cal C}_n$ form a Virasoro algebra of the appropriate
central charge. Indeed, we have 
\be
\begin{split}
[L^{\cal C}_m,L^{\cal C}_n] &= [L^{\cal C}_m,L^{\cal H}_n] -  [L^{\cal C}_m,L^{\cal D}_n]\\
                                     &= [L^{\cal H}_m - L^{\cal D}_m, L^{\cal H}_n]\\
                                     &= (m-n) L^{\cal H}_{m+n} + c^{\cal H} m (m^2-1) \delta_{m,-n} 
                                          - [L^{\cal D}_m, L^{\cal C}_n] - [L^{\cal D}_m,L^{\cal D}_n]\\
                                      &= (m-n) L^{\cal C}_{m+n} + (c^{\cal H} - c^{\cal D}) m (m^2-1) \delta_{m,-n} \ , 
\end{split}
\ee       
where we have used  (\ref{central}), as well as $L^{\cal H}_n = L^{\cal C}_n + L^{\cal D}_n$ (in the penultimate line). This 
is the desired Virasoro algebra of the coset theory. 

In the following we shall always take ${\cal H}$ to be a self-dual theory, i.e., one that has only a single
representation (namely the vacuum representation itself). If we demand modular invariance, such self-dual theories 
exist at $c=24N$ with $N$ integer;\footnote{If we only
demand modular invariance up to a phase then $N$ need not be an integer, but only an integer multiple of $\frac{1}{3}$. The simplest example of such
a theory is the $\mathfrak{e}_8$ theory at level $1$ with $c=8$.}
for $N=1$ corresponding to $c=24$, it is believed that there are precisely $71$ such 
theories  \cite{Schellekens:1992db}, of which all but one have been constructed by now
\cite{Montague:1994yb,Lam}. With the exception of the so-called $e_8^3$ theory, none of them are affine theories.
At $c=24$ all self-dual theories have the character
\be\label{H0ch}
\chi^{\cal H}_0(\tau)  =  J(\tau) + {\cal N} \ , \qquad \hbox{with} \qquad
J(\tau) = j(\tau) -  744 = q^{-1} + 196884 q + \cdots \ , 
\ee
where $q=e^{2\pi i \tau}$, and $j(\tau)$ is the famous modular invariant $j$-function. The integer ${\cal N}$
denotes the number of states with $h=1$, i.e., describes the dimension of the Lie algebra $\mathfrak{g}$ whose affine
Kac-Moody algebra is contained in ${\cal H}$. 

Since the numerator theory ${\cal H}$ contains the denominator chiral algebra, we can decompose ${\cal H}$ in terms
of the irreducible representations of the denominator algebra; on the level of characters, this then leads to
the identity
\be\label{ch0dec}
\chi^{\cal H}_0(\tau) = \chi^{\cal D}_0(\tau) \cdot \chi^{\cal C}_0(\tau) 
+ \sum_{i=1}^{p-1} d_i\, \chi^{\cal D}_i(\tau) \cdot \chi^{\cal C}_i (\tau) \ , 
\ee
where $\chi^{\cal D}_0$ and $\chi^{\cal C}_0$ are the vacuum characters of ${\cal D}$ and ${\cal C}$, respectively, while
$\chi^{\cal D}_i$ with $i=1,\ldots, p-1$ are the remaining $p-1$ irreducible characters of ${\cal D}$. The corresponding branching functions
$\chi_i^{\cal C}$ describe then the characters of the irreducible representations of the coset algebra. The parameter $d_i\in\mathbb{N}$
denote the multiplicities with which these characters appear; since we are only considering specialised characters, two inequivalent
representations may have the same character --- for example, this will be the case for two representations that are 
conjugate to one another ---  and hence non-trivial multiplicities may appear. 

The leading $q^{-1}$ coefficient of (\ref{H0ch}) is reproduced by the first term in (\ref{ch0dec}) since we have 
$c^{\cal D} + c^{\cal C} = c^{\cal H} = 24$. Furthermore, since apart from this term, the left-hand-side only involves 
non-negative integer powers of $q$, we also need that $h_i^{\cal D} + h_i^{\cal C}\equiv n_i \in \mathbb{N}$. (Here $h_i^{\cal D}$
and $h_i^{\cal C}$ are the conformal dimension of the $i$'th character of ${\cal D}$ and ${\cal C}$, respectively.)

If the Lie algebra $\mathfrak{h}$ of ${\cal D}$ is a direct summand of the Lie algebra $\mathfrak{g}$ of ${\cal H}$, 
$\mathfrak{g} = \mathfrak{h} \oplus \mathfrak{k}$, then the coset algebra will contain the affine algebra based on 
$\mathfrak{k}$, and the $q^0$ term of (\ref{ch0dec}) will also arise from the first summand on the right-hand-side; in that 
case we therefore have that $h_i^{\cal D} + h_i^{\cal C} = n_i  \geq 2$. On the other hand, if $\mathfrak{h}$ is not a direct 
summand of $\mathfrak{g}$, then $h_i^{\cal D} + h_i^{\cal C}=n_i =1$ for at least one $i\in\{1,\ldots,p-1\}$.

\section{The modular differential equation}

In the following we will be interested in the modular differential equation that is associated to the coset theory. Let us begin
by briefly reviewing the salient features of this approach to the classification of conformal field theories, see 
\cite{Mathur:1988na,Mathur:1988gt,Naculich:1988xv} for more details. The characters of a rational conformal
field theory satisfy a common modular differential equation. For example, for the case of $2$-character theories
this differential equation is typically of second order, and is then of the form 
\be
\Big(D^2 + \phi_1(\tau)D+\phi_0(\tau)\Big)\, \chi=0 \ , 
\ee
where $D$ is the covariant derivative defined, for example, in \cite{Mathur:1988na}. Since $D$ carries modular weight $2$, 
$\phi_1$ must have modular weight $2$ while $\phi_0$ has modular weight $4$. 

The modular differential equation is further characterised by the number of zeros $\ell\geq 0$ of the associated Wronskian,
which determines the number of poles of the functions $\phi_1$ and $\phi_0$. 
If $\ell=0$, both $\phi_0$ and $\phi_1$ are non-singular, and it 
follows immediately that $\phi_1=0$, while $\phi_0\sim E_4(\tau)$, the familiar Eisenstein series of modular weight $4$. 
The second order equation with $\ell=0$ is then
\be
\bigl(\tD^2 + \mu \,E_4(\tau)\bigr)\, \chi=0 \ , 
\label{leq.0}
\ee
where $\mu$ is a constant. For convenience we have defined $\tD\equiv\frac{D}{2\pi i}$ which simplifies all the 
expressions. The above equation can be written more explicitly in terms of ordinary derivatives as
\be
\left(\tdel^2 -\sfrac{1}{6}E_2(\tau)\, \tdel + \mu \,E_4(\tau) \right) \chi=0 \ , 
\ee
where $\tdel\equiv \frac{\del_\tau}{2\pi i}$, and $E_2$ is the second Eisenstein series (which has a modular anomaly). 

Next consider $\ell=2$. In this case $\phi_0$ and $\phi_1$ can have a pole of maximum degree $\frac{1}{3}$ (the fractional degree 
means that the pole, if it occurs, must be located at $\tau=e^{\frac{i\pi}{3}}$ and counts as a $\frac16$-order pole). Now 
$E_4$ has a double zero at this point, while $E_6$ is non-vanishing. ($E_6$ is the weight-six Eisenstein series, and $E_4$ and 
$E_6$ generate the ring of holomorphic modular forms.) Thus we find $\phi_1\sim \frac{E_6}{E_4}$. On the other hand, 
there is no weight-4 expression that can be made from $E_4$ and $E_6$ that has a single power of $E_4$ in the denominator. 
Hence we must have $\phi_0\sim E_4$, and the $\ell=2$ equation is therefore
\be
\left(\tD^2 + \mu_1\frac{E_6}{E_4}\tD+\mu_2 E_4\right)\chi=0 \ ,
\label{leq.2} 
\ee
where $\mu_1$ and $\mu_2$ are again constants.

For the case of a second order modular differential equation, the situation where $\ell=0$ has been analysed in some detail in 
\cite{Mathur:1988na,Mathur:1988gt,Naculich:1988xv}, while $\ell=2$ has been studied in 
\cite{Naculich:1988xv, Hampapura:2015cea}. In both cases, seven pairs of characters potentially corresponding to unitary 
CFTs were found. The first set is well-understood and consists of the affine theories corresponding to the Lie algebras
$\mathfrak{a}_1$, $\mathfrak{a}_2$,  $\mathfrak{d}_4$, $\mathfrak{g}_2$, $\mathfrak{f}_4$,  $\mathfrak{e}_6$, and
$\mathfrak{e}_7$, all at level $1$. These exhaust all the affine theories for which there are just two characters. The second 
set of characters has not been completely understood in terms of specific CFTs, although in \cite{Hampapura:2015cea} 
some combinations of level-1 affine algebras have been identified that can explain the given central charge and 
first-level degeneracy of the identity character. In Section~\ref{sec:4.2} we shall give an interpretation of these theories in terms of 
the above coset construction.

\subsection{Modular differential equation from cosets}\label{sec:modcos}

Before we can explain the details of this interpretation, let us understand in some generality how the $\ell$ parameters of the 
denominator and coset theories are related to one another. Recall from 
\cite{Mathur:1988na,Mathur:1988gt} that the $\ell$ parameter of the modular differential equation is determined from 
the central charge and the conformal dimensions of the $p$ inequivalent characters via
\be
-\frac{c}{24}+\sum_{i=1}^{p-1}\left(-\frac{c}{24}+h_i\right)=\frac{p(p-1)}{12}-\frac{\ell}{6} \ . 
\ee
Both the denominator theory ${\cal D}$, as well as the coset theory ${\cal C}$ have $p$ inequivalent characters; using that their
central charges add up to $24N$, and that their conformal dimensions add up pairwise to $n_i\in\mathbb{N}$, we find
that the $\ell$ parameter of the coset theory, $\ell^{\,\cal C}$, is determined from the $\ell$ parameter of ${\cal D}$ as 
\be
\ell^{\,\cal C}= p^2 + (6N-1)\, p -6\, \sum_{i=1}^{p-1}n_i-\ell \ . 
\label{genell}
\ee
This relation now allows us to extract some useful information for theories with a small numbers of characters. Suppose that 
$N=1$ and that all the $n_i=2$. Then the above relation specialises to
\be
\ell^{\,\cal C}=(p-3)(p-4)-\ell \ .
\label{spec.3.4}
\ee
So if $p=3$ or $p=4$ the only solutions are $\ell = \ell^{\,\cal C}=0$ (given that both $\ell$ and
$\ell^{\,\cal C}$ have to non-negative). If, on the other hand $p=2$, then 
$\ell^{\,\cal C}=2 - \ell$, and $\ell=2$ leads to $\ell^{\,\cal C}=0$. Below we will find several interesting 
examples of this kind. 

More generally, since both $\ell$ and $\ell^{\,\cal C}$ have to be non-negative, we get interesting constraints on the $n_i$. For 
example, if $N=1$ and $\ell=0$ with $p=3,4$ and all $n_i\geq 2$ (as is the case provided that $\mathfrak{h}$ is a direct summand 
of  $\mathfrak{g}$)  then in fact we must have $n_i=2$ for all $i$.

\section{Interesting Examples}

We are now in the position to explain some of the classification results based on the modular differential equation in terms
of suitable coset constructions. We begin with the
numerological observation about two-character theories with $\ell=0$ that is evident from the table in 
\cite{Mathur:1988na} and the more detailed analysis of \cite{Mathur:1988gt}, and then return to the case of main interest here, 
the two-character theories with $\ell=2$.

\subsection{Two character theories with $\ell=0$}

The two-character theories with $\ell=0$ were classified in 
\cite{Mathur:1988na,Mathur:1988gt}. Excluding the non-unitary cases as well as 
$\mathfrak{e}_8$ level $1$ (a single-character theory) we find that the remaining $7$ theories fall into pairs related by 
$\tc=8-c$ and $\tilh=1-h$. It was shown that for each pair, the fusion rules for the two theories are the same and the 
modular transformation matrices are hermitian conjugates of each other. We can now offer an explanation for this phenomenon. 

We take ${\cal H}$ to be the $\mathfrak{e}_8$ level $1$ theory, i.e., the only one-character theory with $c=8$. As the denominator
theory we take ${\cal D}$ to be any of the affine two-character theories that are contained in $\mathfrak{e}_8$ level $1$, namely 
$\mathfrak{h}= \mathfrak{a}_{1}$, $\mathfrak{a}_{2}$, $\mathfrak{g}_2$, $\mathfrak{d}_4$, $\mathfrak{f}_4$, 
$\mathfrak{e}_6$ or $\mathfrak{e}_7$, all at level $1$. The commutant of $\mathfrak{a}_1$ in $\mathfrak{e}_8$ is $\mathfrak{e}_7$,
and similarly for the pairs $(\mathfrak{a}_2, \mathfrak{e}_6)$ and $(\mathfrak{g}_2,\mathfrak{f}_4)$, while the commutant of 
$\mathfrak{d}_4$ in $\mathfrak{e}_8$ is again $\mathfrak{d}_4$. Since each of these theories has $\ell=0$, it follows 
from eq.~(\ref{genell}) with $N=\frac{1}{3}$, $p=2$ and $\ell=0$ that $n_1=1$ and that $\ell^{\,\cal C}=0$. In particular, the
coset construction therefore relates the $\ell=0$ theories pairwise to one another. Since the modular $S$ matrix of $\mathfrak{e}_8$ 
level $1$ is trivial, the $S$ matrix of the coset is just the hermitian conjugate of that of the denominator theory. This therefore
explains the above numerological observations.

We should note that $n_1=1$ is required in each of these cases since none of the subalgebras is a direct summand of 
$\mathfrak{e}_8$ (given that $\mathfrak{e}_8$ is simple). In particular, the remaining currents of $\mathfrak{e}_8$ that
do not come from the two commuting subalgebras must arise from the $i=1$ term in (\ref{ch0dec}). We have checked
that this works out in each case. Furthermore, since all of these affine subtheories are already 
$2$-character theories, the coset must actually agree with the affine theory. (If the coset was an extension of the affine theory, 
its number of characters would have to be smaller than that of the affine theory, but this is not possible.) 

We should also mention in passing that if we consider $p=3$ affine theories ${\cal D}$, then eq.~(\ref{genell}) with $N=\frac{1}{3}$ 
becomes
\be
\ell^{\, \cal C} = 12 - 6 (n_1 + n_2) - \ell \ , 
\ee
i.e., the only solution with $\ell, \ell^{\,\cal C}\geq 0$ is $\ell^{\,\cal C} = \ell = 0 $ and $n_1=n_2=1$. It would be interesting
to see whether such solutions also exist. (The obvious candidate would be to take ${\cal D}$ the affine theory based on $\mathfrak{a}_3$
at level $1$.)

\subsection{Two-character cosets of meromorphic $c=24$ theories}\label{sec:4.2}

Next we want to explain the relation between the $2$-character theories with $\ell=0$ and $\ell=2$ that was noted in 
\cite{Hampapura:2015cea}, see table~\ref{table-comparison} below for a summary of the salient features.
To understand the relation we start with one of the self-dual Schellekens theories at $c=24$, and consider
the coset with an affine two-character theory ${\cal D}$ with $\ell=0$. (Recall that each of the $\ell=0$ two-character theories 
corresponds to an affine theory based on a Lie algebra at level $1$.) 
We consider the case where the denominator current algebra
is a direct summand of the numerator current algebra, so that $n_1\geq 2$. It then follows from eq.~(\ref{genell}) that $n_1=2$ and 
$\ell^{\,{\cal C}}= 2 - \ell$. Thus this construction will associate to each $\ell=0$ affine two-character theory (whose
Lie algebra appears as a direct summand in one of the Schellekens self-dual theories) a two-character theory with $\ell=2$.

%
%
%

%

\begin{table}[ht]
\centering
\begin{tabular}{|c||c|c|c|c||c|c|c||c|c|}
\hline
 & \multicolumn{4}{c||}{$\ell=0$} & \multicolumn{3}{c||}{$\tll=2$} & \multicolumn{2}{c|}{}\\
\hline
No. & $c$ & $h$ & $m_1$ & Algebra & $\tc$ & $\tilh$ & $\tm_1$ &  $m_1+\tm_1$ & Schellekens No.\\
\hline
1 & 1 & $\frac14$ & 3 &  $\mathfrak{a}_{1}$ & 23 & $\frac74$ & 69 & 72 & $15-21$ \\
\hline
2 &  2 & $\frac13$ & 8 & $\mathfrak{a}_{2}$ & 22 & $\frac53$ & 88 & 96  & $24, 26-28$ \\ 
\hline
3  &  $\frac{14}{5}$ & $\frac25$ & 14 & $\mathfrak{g}_{2}$ & $\frac{106}{5}$ & $\frac85$ & 106  & 120& $32,34$ \\
\hline
4 & 4 & $\frac12$ & 28 & $\mathfrak{d}_{4}$ & 20 & $\frac32$ & 140 & 168 & $42,43$\\
\hline
5 & $\frac{26}{5}$ & $\frac35$ & 52 & $\mathfrak{f}_{4}$ & $\frac{94}{5}$ & $\frac75$ &  188 & 240 & $52,53$\\
\hline
6 & 6 & $\frac23$ & 78 & $\mathfrak{e}_{6}$ & 18 & $\frac43$ & 234  & 312 & $58,59$ \\
\hline
7 & 7 & $\frac34$ & 133 & $\mathfrak{e}_{7}$ & 17 &  $\frac54$ & 323  & 456 & $64,65$\\
\hline
\end{tabular}
\caption{Characters with $\ell=0$ and $\ell=2$. Here $c,\tc$ are the central charges, $h,\tilh$ the conformal dimensions of the primary and $m_1,\tm_1$ the degeneracy of the first excited state in the identity character. All the Lie algebras of the $\ell=0$ theories are at level $1$.}
\label{table-comparison}
\end{table}


In table~\ref{table-comparison} we have listed for each $\ell=0$ theory the Schellekens theories that contain
the corresponding Lie algebra as a direct sumand, see the last column. It is straightforward to work out the 
central charge and the non-trivial conformal dimension  of the corresponding coset theory, and this is 
given in the middle section of the table. These entries then reproduce precisely the findings of 
\cite{Naculich:1988xv,Hampapura:2015cea}. The chiral algebra of the coset theory contains the affine algebra that 
is obtained from the direct sum of affine algebras of the numerator by deleting the affine algebra of the denominator. 
However, the full chiral algebra of the coset is not just generated by these currents. Given that there are 
different self-dual $c=24$ theories from which we may start (that differ by the affine symmetry algebra they contain), 
it is clear that there are also different $\ell=2$ coset theories (that differ again by their affine subalgebra) with the 
same pair of characters.

On the level of the chararacters, the construction implies that we have the relation 
\be
J(\tau) + {\cal N} = \chi_0(\tau) \tchi_0(\tau) + \chi_1(\tau) \tchi_1(\tau) \ , 
\label{branchfnltwo}
\ee
where ${\cal N}= m_1 + \tilde{m}_1$ (since $n_1=2$). The characters of the theories in the table are known exactly as 
hypergeometric functions \cite{Naculich:1988xv},
\be
\begin{split}
\chi_0 &=  j^{\frac{c}{24}}~{}_2 F_1\!\!\left(-\shalf\big(h-\sfrac16\big), -\shalf\big(h-\sfrac56\big);1-h;\sfrac{1728}{j}\right) \\
\chi_1 &=  \sqrt{m}~j^{\frac{c}{24}-h}~{}_2 F_1\!\!\left(\shalf\big(h+\sfrac16\big), \shalf\big(h+\sfrac56\big);1+h;\sfrac{1728}{j}\right) \\ 
\tchi_0 &=   j^{\frac{\tc}{24}}~{}_2 F_1\!\!\left(-\shalf\big(\tilh+\sfrac16\big), -\shalf\big(\tilh-\sfrac76\big);1-\tilh;\sfrac{1728}{j}\right) \\
\tchi_1 &=  \sqrt{\tm}~j^{\frac{\tc}{24}-\tilh}~{}_2 F_1\!\!\left(\shalf\big(\tilh-\sfrac16\big), \shalf\big(\tilh+\sfrac76\big);1+\tilh;\sfrac{1728}{j}\right) \ , 
\end{split}
\label{charexp}
\ee
where
\be
\begin{split}
\sqrt{m} & = (1728)^h\, \left(\frac{\sin \frac{\pi}{2}\left(\sfrac16-h\right)\sin \frac{\pi}{2}\left(\sfrac56 -h\right)}
{\sin \frac{\pi}{2}\left(\sfrac16+h\right)\sin \frac{\pi}{2}\left(\sfrac56 +h\right)}\right)^\shalf
\frac{\Gamma(1-h) \Gamma\Big(\half\big(\frac{11}{6}+h\big)\Big) \Gamma\Big(\half\big(\frac76 + h\big)\Big)}
{\Gamma(1+h) \Gamma\Big(\half\big(\frac{11}{6}-h\big)\Big) \Gamma\Big(\half\big(\frac76 - h\big)\Big)}\\
\sqrt{\tm} & = (1728)^\tilh\, \left(\frac{\sin \frac{\pi}{2}\left(\sfrac16+\tilh\right)\sin \frac{\pi}{2}\left(\sfrac76 -\tilh\right)}
{\sin \frac{\pi}{2}\left(\sfrac16-\tilh\right)\sin \frac{\pi}{2}\left(\sfrac76 +\tilh\right)}\right)^\shalf
\frac{\Gamma(1-\tilh) \Gamma\Big(\half\big(\frac{13}{6}+\tilh\big)\Big) \Gamma\Big(\half\big(\frac56 + \tilh\big)\Big)}
{\Gamma(1+\tilh) \Gamma\Big(\half\big(\frac{13}{6}-\tilh\big)\Big) \Gamma\Big(\half\big(\frac56 - \tilh\big)\Big)} \ .
\end{split}
\ee
For the specific values of $h$ corresponding to the known $\ell=0$ CFTs (the left half of the table), one can easily calculate $m$ and $\tm$. 
These correspond to the degeneracies of the ground state of the nontrivial primary, along with a factor to account for the possible 
multiplicity of primaries with the same character. When such multiplicities are absent, both $m$ and $\tm$ are integers. Otherwise, 
they differ from an integer by a factor of $\sqrt2$ or $\sqrt3$ (these are the only possible multiplicities encountered, the first coming 
from complex conjugation and the second from triality of $\mathfrak{d}_4$). The relevant results can be found in Table 1 of 
\cite{Mathur:1988na} and Table 2 of \cite{Naculich:1988xv}, respectively. One sees that the product $\sqrt{mm'}$ is in any case an integer. 

We have checked that the relevant character identities (\ref{branchfnltwo}) indeed work out in every case. In fact, this 
is a consequence of a relation of hypergeometric functions,
\begin{eqnarray}
\label{sphypid}
& & {}_2 F_1(r,r+\tfrac{1}{3};2r+\tfrac{5}{6};x) \, {}_2 F_1(-r-1,-r-\tfrac{1}{3};-2r-\tfrac{5}{6};x)   \\
& & \quad +\, M x^2  \, {}_2 F_1 (-r+\tfrac{1}{6},-r+\tfrac{1}{2};-2r+\tfrac{7}{6};x)\,  {}_2 F_1(r+\tfrac{5}{6},r+\tfrac{3}{2};2r+\tfrac{17}{6};x)     
=   1 - \frac{2 (3 r+1)} { (12 r+5) } x   \ ,     \nonumber
\end{eqnarray}
where
\be
M=\frac{216 (2 r+1) (r+1) (3r+1) r } {(12r+11) (12r+5)^2 (12r-1) } = (1728)^{-2} \, \sqrt{m \tilde{m}} \ .
\ee
This in turn is a special case of the hypergeometric identity
\begin{eqnarray}
& &{}_2 F_1(a,b;c;x) \, {}_2 F_1(-a-1,-b,-c;x)  \nonumber \\
&&\quad +\,\frac{ab(a+1)(b-c)x^2}{c^2(1-c^2)}{}_2 F_1(a-c+1,b-c+1;2-c;x) \, {}_2 F_1(-a+c,1-b+c;2+c;x) \nonumber \\
&&\quad\quad +\,\frac{b}{c}x=1 \ . \label{genhypid}
\end{eqnarray}
This can be proven by noting that the hypergeometric functions are meromorphic in the parameter $c$, with simple poles 
at all non-positive integers; at the poles the behaviour is
\be
\lim_{c\to -n} \frac{{}_2F_1(a,b;c;x)}{\Gamma(c)}=\frac{(a)_{n+1}(b)_{n+1}x^{n+1}}{(n+1)!}{}_2F_1(a+n+1,b+n+1;n+2;x) \ .
\ee
Now each of the first two terms in eq.~(\ref{genhypid}) has a single pole for all $c\in \IZ$, $c\ne0$. Using the above identity, it is easily 
verified that the residues at these poles cancel between the two terms. As $c\to 0$ one has both double and simple poles from 
the first two terms, and a simple pole from the third term. Again one can check that the residues cancel. From the cancellation 
of all poles, it follows that the LHS is constant in $c$. Next, choosing $c=b$ one finds that this constant is equal to $1$, 
independent of $a,b$ and $x$. 

\eref{genhypid} implies eq.~(\ref{sphypid}) via the substitution $a=r$, $b=r+\frac13$, $c=2r+\frac56$. In turn, this implies the 
character identities of eq.~(\ref{branchfnltwo}) upon writing  
\be
x = \frac{1728}{j} \ , \quad r= - \frac{1}{2} (h-\tfrac{1}{6})   \ .
\ee
Furthermore, the constant term equals
\be
{\cal N} = 744 - 1728 \frac{2 (3 r+1)} { (12 r+5) } \ ;
\ee
this will obviously only be an integer if $r$ is correctly chosen (such as is the case for the entries of the table). 

%
%

\subsection{Three-character and Four-character cosets of $c=24$}

The above observations provide a practical way to generate many conformal field theories with a small number of characters, 
starting from affine theories with the same number of characters. In order to exhibit this, let us consider 
affine 3-character or 4-character theories ${\cal D}$ with $\ell=0$ whose algebra is a direct summand of any of the $71$ self-dual
$c=24$ theories ${\cal H}$ of \cite{Schellekens:1992db}. Because of the discussion at the end of section~\ref{sec:modcos}, we know 
that the coset theory will then also have $\ell^{\,{\cal C}}=0$. As we shall see, this will give rise to interesting solutions of the
modular differential equation of \cite{Mathur:1988gt} that satisfy all the required integrality conditions. In this  subsection we shall describe all
the examples that arise in this manner.


As a first example, let us take ${\cal D}$ the $\mathfrak{a}_3$ theory at level $1$. This theory has three characters and
leads to a modular differential equation with $\ell=0$. Its central charge is $c=3$, and the conformal dimensions are 
$(h_1,h_2)=(\frac{3}{8}, \half)$. This ${\cal D}$ theory is contained, as a direct summand, in the self-dual theory number $30$
of Schellekens' list \cite{Schellekens:1992db}; the latter theory has ${\cal N}=120$ since its current algebra is $\mathfrak{a}_3^{\oplus 8}$. 
(The dimension of $\mathfrak{a}_3$ is $15$.) Given the discussion above, the coset theory will have $\tilde{c}=21$, and possess the current algebra $\mathfrak{a}_3^{\oplus 7}$
of dimension $105$. From the analysis of section~\ref{sec:modcos} we also know that the $n_i=2$, so its conformal dimensions must 
be $(\tilde{h}_1,\tilde{h}_2)=(\frac{13}{8},\frac{3}{2})$. Furthermore, it will have $\ell^{\,{\cal C}}=0$. This data is then sufficient
to fix the differential equation satisfied by the coset theory completely, and as a consequence allows one to calculate the characters
(as solutions of the modular differential equation). We have checked that the degeneracy at the first level in the identity character
comes out to be $105$. We have also computed the Fourier coefficients of the three characters to very
high orders and verified that they are indeed positive integers for the identity character, and rational numbers (when normalised 
so the ground state is unity) for the other characters. Among other things this demonstrates that the modular differential equation
approach is a useful method for the determination of the branching functions of the coset. 

All the remaining cases work out equally nicely, and our results for the three-character
examples are summarised in table~\ref{table-3characters}. Our coset theories all contain
an affine algebra --- the algebra generated by the remaining summands of the numerator theory 
that are not divided out by taking the coset --- but none of them is just generated by this affine algebra;
in particular, all of these theories therefore seem to be new.

The situation for the four-character examples is essentially identical, but provides a much smaller list of examples
that are summarised in table~\ref{table-4characters}. Again, the resulting four-character theories are not just generated
by the affine currents and seem to be new.


\begin{table}[ht]
	\centering
	\begin{tabular}{|c||c|c|c|c|c| |c|c|c|c||c|c|}
		\hline
		& \multicolumn{5}{c||}{${\cal D}$} & \multicolumn{4}{c||}{${\cal C}$} & \multicolumn{2}{c|}{}\\
		\hline
		No. & $c$ & $h_1$ & $h_2$ &$m_1$ & Algebra & $\tc$ & $\tilh_1$ & $\tilh_2$ & $\tm_1$ &  $m_1+\tm_1$ & Schellekens No.\\
		\hline
		1  &  $\frac{3}{2}$ &  $\frac{3}{16}$ & $\frac12$& 3 & $\mathfrak{a}_{1,2}$ & $\frac{45}{2}$ & $\frac{29}{16}$ & $\frac32$ & 45 &  48 &  $5,7,8,10$ \\
		\hline
		2 & $\frac52$ & $\frac{5}{16}$ & $\frac12$ & 10 & $\mathfrak{c}_{2,1}$ & $\frac{43}{2}$ & $\frac{27}{16}$ & $\frac32$ & 86 & 96 & $25,26,28$\\
		\hline
		3 & 3 & $\frac38$ & $\frac12$ & 15 &  $\mathfrak{a}_{3,1}$ & 21 & $\frac{13}{8}$ & $\frac32$ & 105 & 120 & $30,31,33-35$ \\
		\hline
		4 & $\frac{7}{2}$ & $\frac{7}{16}$ & $\frac12$& 21 & $\mathfrak{b}_{3,1}$ & $\frac{41}{2}$ & $\frac{25}{16}$ & $\frac32$ & 123 & 144 & $39,40$\\
		\hline
		5 &  4 & $\frac25$ & $\frac35$ & 24 & $\mathfrak{a}_{4,1}$ & 20 & $\frac85$ & $\frac75$ & 120 & 144 & $37,40$ \\ 
		\hline
	
		6 & $\frac92$ & $\frac{9}{16}$ & $\frac12$& 36 & $\mathfrak{b}_{4,1}$ & $\frac{39}{2}$ & $\frac{23}{16}$ & $\frac32$ & 156 & 192 & $47,48$ \\
		\hline
		7 & $5$ & $\frac{5}{8}$ & $\frac12$& 45 & $\mathfrak{d}_{5,1}$ & $19$ & $\frac{11}{8}$ & $\frac32$ & 171 & 216 & $49$\\
		\hline
		8 & $\frac{11}{2}$ & $\frac{11}{16}$ & $\frac12$& 55 & $\mathfrak{b}_{5,1}$ & $\frac{37}{2}$ & $\frac{21}{16}$ & $\frac32$ & 185 & 240 & $53$\\
		\hline
		9 & $6$ & $\frac{3}{4}$ & $\frac12$& 66 & $\mathfrak{d}_{6,1}$ & $18$ & $\frac{5}{4}$ & $\frac32$ & 198 & 264 & $54,55$\\
		\hline
		10 & $\frac{13}{2}$ & $\frac{13}{16}$ & $\frac12$& 78 & $\mathfrak{b}_{6,1}$ & $\frac{35}{2}$ & $\frac{19}{16}$ & $\frac32$ & 210 & 288 & $56$\\
		\hline
		11 & $7$ & $\frac{7}{8}$ & $\frac12$& 91 & $\mathfrak{d}_{7,1}$ & $17$ & $\frac{9}{8}$ & $\frac32$ & 221 & 312 & $59$\\
		\hline
        12 & $\frac{17}{2}$ & $\frac{17}{16}$ & $\frac12$& 136 & $\mathfrak{b}_{8,1}$ & $\frac{31}{2}$ & $\frac{15}{16}$ & $\frac32$ & 248 & 384 & $62$\\
		 \hline
	    13 & $\frac{31}{2}$ & $\frac{15}{16}$ & $\frac32$& 248 & $\mathfrak{e}_{8,2}$ & $\frac{17}{2}$ & $\frac{17}{16}$ & $\frac12$ & 136 & 384 & $62$\\
		  \hline
		14 & $9$ & $\frac{9}{8}$ & $\frac12$& 153 & $\mathfrak{d}_{9,1}$ & $15$ & $\frac{7}{8}$ & $\frac32$ & 255 & 408 & $63$\\
		 \hline
		15 & $10$ & $\frac{7}{8}$ & $\frac12$& 190 & $\mathfrak{d}_{10,1}$ & $14$ & $\frac{9}{8}$ & $\frac32$ & 266 & 456 & $64$\\
		 \hline
		
	\end{tabular}
	\caption{Three-character theories with $\ell=0$. Here $c,\tc$ are the central charges, $h_1,h_2,\tilh_1,\tilh_2$ 
	the conformal dimensions of the primaries and $m_1,\tm_1$ the degeneracy of the first excited state in the identity character.}
	\label{table-3characters}
\end{table}

\begin{table}[ht]
	\centering
	\begin{tabular}{|c||c|c|c|c|c|c| |c|c|c|c|c||c|c|}
		\hline
		& \multicolumn{6}{c||}{${\cal D}$} & \multicolumn{5}{c||}{${\cal C}$} & \multicolumn{2}{c|}{}\\
		\hline
		No. & $c$ & $h_1$ & $h_2$ & $h_3$ & $m_1$ & Algebra & $\tc$ & $\tilh_1$ & $\tilh_2$ & $\tilh_3$ & $\tm_1$ &  $m_1+\tm_1$ & Schellekens No.\\
		\hline
		1 & $\frac{16}{5}$ & $\frac{4}{15}$ & $\frac{2}{3}$ & $\frac{3}{5}$ & 8 & $\mathfrak{a}_{2,2}$ & $\frac{104}{5}$ & $\frac{26}{15}$ & $\frac43$ & $\frac75$  & 52 & 60 & $13,14$\\
		
		\hline
		2 & $\frac{14}{3}$ & $\frac13$ & $\frac23$& $\frac79$ & 14 & $\mathfrak{g}_{2,2}$ & $\frac{58}{3}$ & $\frac{4}{3}$ & $\frac53$& $\frac{11}{9}$ & 58 & 72 & $21$\\
		\hline
		
		3 & $\frac{21}{5}$ & $\frac{7}{20}$ & $\frac35$ & $\frac34 $& 21 &  $\mathfrak{c}_{3,1}$ & $\frac{99}{5}$ & $\frac{33}{20}$ & $\frac75$ & $\frac54$ & 99 & 120 & $33$ \\
		\hline
		
	    4  &  $5$ &  $\frac{5}{12}$ & $\frac{2}{3}$ & $\frac{3}{4}$ & 35 & $\mathfrak{a}_{5,1}$ & $19$ & $\frac{19}{12}$ & $\frac43$ & $\frac54$ &  133 & 168 &  $43-45$ \\
		\hline
	
		\end{tabular}
			\caption{Four-character theories with $\ell=0$. Here $c,\tc$ are the central charges, $h_1,h_2,h_3,\tilh_1,\tilh_2,\tilh_3$ 
			the conformal dimensions of the primaries and $m_1,\tm_1$ the degeneracy of the first excited state in the identity character.}
			\label{table-4characters}
		\end{table}

\section{Conclusions}

In this paper we have constructed interesting examples of conformal field theories by taking cosets of the self-dual
$c=24$ meromorphic theories (that only have a single character) by some affine subtheory with a small number of characters. 
The situation is particularly simple if the affine algebra of the denominator is a direct summand of that of the numerator,
and we have in this way constructed the two-character RCFTs with $\ell=2$ that were previously predicted on the level
of the modular differential equation. We have also used the same idea to provide new interesting examples of 
$\ell=0$ RCFTs with three and four characters.  

%

Many generalisations of our construction are possible, for example one can consider cosets of general meromorphic CFTs 
with central charge $c=24k$ where $k> 1$. Although the number of examples rapidly becomes enormous, it would be 
interesting to investigate whether this provides some insight into the classification problem for RCFTs.

\section*{Acknowledgements}

We thank J\"urgen Fuchs and Roberto Volpato for discussions and correspondences. 
MRG thanks the ITP at the Chinese Academy of Sciences for hospitality during the final
stages of this work, while SM acknowledges the hospitality of ETH Z\"urich, and in particular
of the Pauli Center, during the initial stages of this work. The work of MRG is 
(partly) supported by the NCCR SwissMAP, funded by the Swiss National Science Foundation. The work of 
HRH is supported by an INSPIRE Scholarship, DST, Government of India, and that of SM by a J.C.~Bose Fellowship, 
DST, Government of India. We thank the people of India for their generous support for the basic sciences.



\bibliography{coset}

\end{document}